\documentclass{article}

\usepackage{arxiv}
\usepackage[utf8]{inputenc} 
\usepackage[T1]{fontenc}    
\usepackage[nolist]{acronym}  
\usepackage{hyperref}   
\usepackage{graphicx}
\usepackage{csquotes}
\usepackage{censor}
\usepackage{rotating}
\usepackage{lscape}
\usepackage{array}
\usepackage{booktabs}
\usepackage{threeparttable}
\usepackage{threeparttablex}
\usepackage{longtable}
\usepackage{microtype}
\usepackage{xcolor}
\usepackage{doi}
\definecolor{codegray}{gray}{0.95}
\usepackage{siunitx}

\usepackage
[backend=biber,
style=authoryear-comp,
sorting=ynt]{biblatex}
\usepackage{hyperref}
\hypersetup{colorlinks=true, linkcolor=blue, urlcolor=cyan}

\title{Mapping Public Perception of Artificial Intelligence: Expectations, Risk-Benefit Tradeoffs, and Value  As Determinants for Societal Acceptance}

\renewcommand{\shorttitle}{Mapping Public Perception of Artificial Intelligence}

\author{
\href{https://orcid.org/0000-0003-2837-5181}{\includegraphics[scale=0.06]{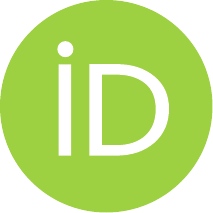}\hspace{1mm}Philipp Brauner} \\
	Communication Science\\
	RWTH Aachen University\\
	52056 Aachen, Germany\\
	\texttt{brauner@comm.rwth-aachen.de} \\
\And
\href{https://orcid.org/0000-0001-7054-8520}{\includegraphics[scale=0.06]{orcid.pdf}\hspace{1mm}Felix Glawe} \\
	Communication Science\\
	RWTH Aachen University\\
	52056 Aachen, Germany\\
	\texttt{brauner@comm.rwth-aachen.de}
\And
\href{https://orcid.org/0000-0001-5065-1619}{\includegraphics[scale=0.06]{orcid.pdf}\hspace{1mm}Gian Luca Liehner} \\
	Communication Science\\
	RWTH Aachen University\\
	52056 Aachen, Germany\\
	\texttt{brauner@comm.rwth-aachen.de}
\And
\href{https://orcid.org/0000-0002-9030-6999}{\includegraphics[scale=0.06]{orcid.pdf}\hspace{1mm}Luisa Vervier} \\
	Communication Science\\
	RWTH Aachen University\\
	52056 Aachen, Germany\\
	\texttt{brauner@comm.rwth-aachen.de}
\And
\href{https://orcid.org/0000-0002-6105-4729}{\includegraphics[scale=0.06]{orcid.pdf}\hspace{1mm}Martina Ziefle} \\
	Communication Science\\
	RWTH Aachen University\\
	52056 Aachen, Germany\\
	\texttt{ziefle@comm.rwth-aachen.de}
}

\begin{document}

\begin{acronym}
	\acro{ai}[AI]{Artificial Intelligence}
	\acro{ml}[ML]{Machine Learning}
	\acro{dl}[DL]{Deep Learning}
	\acro{llm}[LLM]{Large Language Model}
	\acroplural{llm}[LLMs]{large language models}
	\acro{ca}[CA]{Conjoint Analysis}
	\acro{tam}[TAM]{Technology Acceptance Model}
	\acro{vam}[VAM]{Value-Based Acceptance Model}
\end{acronym}

\maketitle

\begin{abstract}
Understanding public perception of artificial intelligence (AI) and the tradeoffs between potential risks and benefits is crucial, as these perceptions might shape policy decisions, influence innovation trajectories for successful market strategies, and determine individual and societal acceptance of AI technologies.
Using a representative sample of 1100 participants from Germany, this study examines mental models of AI.
Participants quantitatively evaluated 71 statements about AI’s future capabilities (e.g., autonomous driving, medical care, art, politics, warfare, and societal divides), assessing the expected likelihood of occurrence, perceived risks, benefits, and overall value.
We present rankings of these projections alongside visual mappings illustrating public risk-benefit tradeoffs.
While many scenarios were deemed likely, participants often associated them with high risks, limited benefits, and low overall value.
Across all scenarios, 96.4\% ($r^2=96.4\%$) of the variance in value assessment can be explained by perceived risks ($\beta=-.504$) and perceived benefits ($\beta=+.710$), with no significant relation to expected likelihood.
Demographics and personality traits influenced perceptions of risks, benefits, and overall evaluations, underscoring the importance of increasing AI literacy and tailoring public information to diverse user needs.
These findings provide actionable insights for researchers, developers, and policymakers by highlighting critical public concerns and individual factors essential to align AI development with individual values.
\end{abstract}

\keywords{AI Ethics, cognitive maps, technology acceptance, psychometric paradigm, technology assessment, mental models, artificial intelligence}

\acresetall
\section{Introduction}\label{sec:introduction}

The rapid advancements in \ac{ai} and \ac{dl} in general and \acp{llm} in particular, has ignited widespread interest and concern across multiple domains.
These technologies become increasingly integrated into almost all sectors ranging from education \parencite{Chen2020} and healthcare \parencite{Amunts2024} to journalism \parencite{Diakopoulos2019}, forestry and farming \parencite{Holzinger2024}, as well as production and manufacturing \parencite{Brauner2021}.
They offer benefits in terms of efficiency, convenience, and innovation \parencite{Bouschery2023}, but also pose significant risks in terms of privacy infringement, job displacement \parencite{Acemoglu2017}, and ethical dilemmas for individuals, organizations, and society as a whole \parencite{Awad2018}.

Although AI’s origins stretch back several decades \parencite{McCarthy2006,Hopfield1982,Rumelhart1986}, its development has accelerated significantly in recent years, driven by advancements in computing power, greater availability of digital data \parencite{Deng2009}, enhanced algorithms, and a considerable surge in funding \parencite{Lecun2015,statistaAI2022}.
The expectations surrounding \ac{ai} appear to be split: while some researchers and consumers regard AI as a transformative tool that will enhance our lives \parencite{Brynjolfsson2014,Makridakis2017,Bouschery2023}, others voice concerns about its ethical implications and associated risks \parencite{Cath2018,Bostrom2003,Crawford2021}.

For decades, it has been recognized that computers and algorithms are not value-neutral but inherently embody values and potential biases \autocite{Friedman1996, Nissenbaum2001}.
Through the  currently has a unique yet ambiguous position within the human cognitive and social landscape \parencite{budish2021ai,garcia2024keeping}, prompting a critical evaluation of how people perceive AI and the broader societal implications they associate with its adoption.
This underscores the need for critical scrutiny in the design and deployment of technology, as the embedded values can influence decisions and outcomes in ways that may perpetuate inequality or reinforce existing biases. 
Further, the deployment and successful use of technologies, such as \ac{ai}, can be accelerated by higher acceptance or hindered by perceived obstacles \parencite{young2021patient}.

There is a shared concern about the ethical and societal impacts of AI, with a need for careful design and forward-looking research policies to avoid setbacks \parencite{Gursoy2023}.
Hence, understanding public perception of artificial intelligence, particularly how people balance its perceived benefits and risks, is essential as these views shape policy decisions, influence innovation trajectories, and determine the individual and societal acceptance of AI technologies \parencite{Sadek2024}.
This article examines how the public evaluates AI’s potential impact and future capabilities, exploring the tradeoffs they consider between AI’s utilities and associated risks.
The results are analysed both at the level of individual differences, meaning how personality factors influence the perception of \ac{ai}, as well as at the technological level, meaning how the risk and benefit perception of the different projections shape the overall evaluation of \ac{ai}.
We further illustrate the risk and benefit tradeoff by placing the evaluations on visual maps.
Overall, this article helps to identify topics where risk and benefit expectations are aligned, as well as topics with greater differences and potential for conflict in terms of public perception and acceptance.

The structure of the article is as follows:
Section \ref{section:relatedwork} gives an overview on the public perception of \ac{ai} as well as an overview on technology and risk perception.
Section \ref{sec:method} presents the concept of using micro scenarios as the basis for our empirical approach, the survey, and the sample.
In Section \ref{ref:results} we present the results of the study starting with the overall perception of the various \ac{ai}-related statements and their visual mapping, followed by an analysis of the influence of the individual differences.
Section \ref{sec:discussion} discusses the findings and their implications, as well as the limitations of the study.
Lastly, Section \ref{sec:conclusion} suggests policy implications and future research.

\section{Related Work}\label{section:relatedwork}

We first provide a brief overview on risk perception and the psychometric model for measuring subjective risk.
Thenm we then give an overview on studies on the public opinion and perception of \ac{ai}.
 
\subsection{Risk Perception and the Psychometric Model}


Across many domains, individual risk and benefit perceptions influence attitudes, usage intentions, or actual behaviors \parencite{Witte2000,Hoffmann2015,Huang2020}.
Risk can be conceptualised from two perspectives \parencite{Aven2009,Fischhoff1978}:
On one hand, risk can be modeled as the expected utilities of negative events, their potential consequences, and related to individual responses \parencite{Kahneman1984}.
However, a key challenge is to accurately model the probabilities and consequences, making it difficult to calculate expected utilities and link them to individual responses.
On the other hand, risk can be viewed as events and their perceived consequences.
According to this psychometric model of risk perception, an individual’s perceived threat from these consequences can be measured, for example, by means of rating scales \parencite{Slovic1979,Fischhoff1978, Slovic1986}.
For example, it was used to explore the role of risk perception in areas such as gene technology \parencite{Connor2010}, genetically modified food \parencite{Verdurme2003}, nuclear energy \parencite{Slovic2000}, general climate change \parencite{Pidgeon2011}, and Carbon Capture and Utilization technologies \parencite{Arning2020}.
It represents a framework for understanding how individuals perceive and balance risks and benefits and link these weightings with related dispositions and properties of the technologies.
Therefore, it is particularly useful for studying perceptions of emerging technologies or technologies whose implications are not easily understood by laypeople.
This is even more important in the early phases of technology development, to assess potential social and societal consequences through the lens of laypeople, even though they might have a limited and fragmented understanding of the impacts of AI and the resulting changes in society, the job market, or other areas.
Independently from the specific context, a commonly observed pattern is an inverse relationship between perceived risks and perceived benefits \parencites{Alhakami1994}:
When a technology is perceived as risky, its benefits are often viewed as less significant, whereas technologies perceived as safer tend to be associated with higher perceived benefits, and vice versa.

\subsection{Perception of Artificial Intelligence}

Studies reveal a complex landscape of attitudes and perceptions towards AI that vary by time, context, and individual.
But as \ac{ai} spans numerous tasks and domains, it is challenging to provide a comprehensive overview of the research on public perceptions of \ac{ai}, particularly in relation to the associated risk-benefit tradeoffs.
Especially the rapid and unprecedented adoption of ChatGPT following its public release in 2022  \parencite{Hu2023chatgpt} spurred academic interest on perception of AI and much of this research remains to be consolidated.
Henceforth, the following section offers a brief and necessarily incomplete overview of this rapidly evolving field.

\subsubsection{General AI Perception}

\parencite{Fast2017} conducted a media analysis of AI coverage in the New York Times spanning three decades.
They found a growing public interest in \ac{ai} after 2009, marked by both optimism and concern, with AI generally receiving more positive than negative coverage. 
Yet, recent years have seen rising concerns about control loss and ethical dilemmas, contrasting with optimism, particularly regarding AI's potential for healthcare.
People often hold inflated expectations about AI’s potential, driven in part by optimistic portrayals in news and entertainment media \parencite{Fast2017}.

Recently, \textcite{Sanguinetti2024}
examined how news outlets convey \enquote{AI anxiety} by depicting AI as an autonomous, opaque entity independent of human control.
They derived an AI anxiety index and analyzed headlines across major newspapers before and after ChatGPT’s launch.
Their findings indicate that ChatGPT’s introduction not only increased AI-related coverage but also intensified negative sentiments, with regional media driving the heightened AI anxiety index.

Research indicates that public awareness of AI technologies is still generally limited, with many people struggling to differentiate between specific forms like machine learning, robotics, and automation.
A survey by Ipsos found that the general population often lacks a nuanced understanding of AI’s technical achievements and limitations \parencite{IpsosAI2022}.
Pew Research found that only a fraction of Americans could correctly identify AI in everyday scenarios, highlighting a general lack of clarity about AI’s scope and capabilities \parencite{pew2023ai}.
This limited awareness contributes to misconceptions and oversimplified views of AI’s impact and applications, potentially impeding informed public discourse on AI’s ethical and societal implications.
The Alan Turing Institute also highlighted that public understanding varies significantly depending on education level and context, with frequent concerns about automation and robotics in particular, such as in employment and security applications \parencite{turing2023}.

Public discourse often includes both, unfocused and generic fears on the one hand as well as high expectations about AI on the other, particularly around the concept of artificial general intelligence (AGI), which still remains largely fictional \parencite{Jungherr2023}.
\textcite{Cave2019} investigated prevalent narratives about AI using a sample from the UK, identifying eight primary themes—four optimistic and four pessimistic.
Their findings suggest that perceptions of AI’s impact are often tinged with anxiety, with only two of the narratives viewing benefits as outweighing concerns (such as the idea that AI could make life easier).
Additionally, participants expressed a sense of powerlessness over AI development, viewing it as largely driven by government and corporate interests.
About half of the respondents were able to provide plausible definitions of AI, while 25\% associated AI primarily with robots.

Further research indicates that public concern over AI’s ethical use is on the rise, particularly as awareness of biased algorithms and discriminatory outcomes has grown \autocite{ONeil2016}.
Scholars argue that transparency, accountability, and fairness are key factors in building public trust in AI systems \parencite{Floridi2018,Binns2018}.

In an earlier study, we examined the expectations (is the AI projection likely or not likely to occur) and sentiment (is the use of AI negative or positive) of laypersons towards AI-related scenarios by utilizing a younger convenience sample \parencite{Brauner2023}.
The findings suggest greater disparities in the perceptions of these subjects:
Of particular concern was the expectation of cyber security threats, a factor deemed both highly likely and least favorable by the participants.

\subsubsection{Context dependency of AI perception}

Context-wise, studies on the (public) perception of domain-specific uses of \ac{ai} are relatively common.
Perceptions of AI risks and benefits vary across different domains such as healthcare, education, and creative arts, with healthcare often seen as more beneficial \parencite{Novozhilova2024}.
\textcite{Alessandro2024} suggest that \ac{ai}'s perceived social risk and value are inversely related; higher perceived risks lead to lower attributed social value.
In this study, medical AI applications were perceived as particularly risky.
\textcite{Gao2020} studied the perception of \ac{ai} in medical care in China through content analysis of social media posts.
Key concerns associated with a negative evaluation were the immaturity of the technology and distrust in the related companies. Further, in the majority of the posts replacing human doctors being replaced with \ac{ai} was expected.

With an experimental approach \textcite{Liehner2021} studied the willingness to delegate morally sensitive tasks to automated AI-agents and found that context and reliability (i.e., risk of an error) of the automation shapes the perception of and trust in AI.
A meta-analysis on risk perception of narrow AI (AI for specific tasks) found that key influences that mitigate risk perception are familiarity, trust, whereas privacy concerns exacerbate the perceived risks of a technology \parencite{Krieger2024}.
\textcite{Araujo2020} explored how individual differences related to the perceptions of automated decision-making by AI and how AI perception differs by context (media, (public) health, and judicial).
People were concerned about risks and had mixed opinions about fairness and usefulness of automated decision-making at a societal level, although  AI-based decisions were evaluated on par or even better than human experts for specific decisions.
In the study, AI knowledge had a positive influence on perceived benefits and fairness of AI, whereas privacy concerns were linked to the perceived risks.

\subsubsection{AI Perception and Individual Differences}

Individual differences, such as demographics (gender, age) but also experience with and knowledge of AI influence the perception of AI \parencite{Yigitcanlar2022}.
For instance, people with higher technological competence and AI familiarity tend to trust AI more \parencite{Novozhilova2024,Crockett2020}.
\textcite{Kaya2022} investigated general attitudes towards artificial intelligence and the influence of personality traits with a Turkish sample.
Again, computer use and knowledge about AI had a positive influence on attitude towards AI.
Agreeableness, AI learning anxiety, and AI configuration anxiety had a negative influence on the attitudes towards AI.


\textcite{deWinter2024} studied the relationship between personality factors, performance expectations, and intention to use and actual use of ChatGPT.
Perceived effectiveness and concerns correlated with ChatGPT use frequency. Further, intention to use was linked to the personality trait  Machiavellianism (i.e., use of manipulation tactics).

\textcite{Kelley2021} conducted a study on public opinion regarding \ac{ai}, surveying over 10,000 participants from the eight countries Australia, Canada, the USA, South Korea, France, Brazil, India, and Nigeria.
The study examined the anticipated societal impact of AI and participants’ attitudes toward it, using four key descriptors: exciting, useful, worrying, and futuristic.
In developed countries, such as the USA, Canada, and Australia, respondents predominantly expressed concerns about AI, coupled with futuristic expectations.
In contrast, developing countries like India, Brazil, and Nigeria exhibited a greater sense of excitement about AI’s potential.
South Korea stood out for its focus on AI’s usefulness and future applications, reflecting its advanced technological landscape. Across all regions, there was a broad consensus that AI would have a significant societal impact, though the exact implications were still uncertain.

Studying the perceptions of participants from China on a few selected \ac{ai}-based technologies \textcite{Cui2019} found an overall positive attitude towards AI and that the perceived benefits of AI outweigh the perceived risks.
Notably, media usage was solely correlated with positive perceptions of AI's advantages, potentially influenced by government-controlled media that presents \ac{ai} in a favourable manner.
Further, individuals with a high level of personal relevance exhibited reduced susceptibility to media influence, fostering a more critical stance towards AI.

A comparative media analysis explored the similarities and differences in coverage of the historic chess match between Lee Sedol and AlphaGo \parencite{Curran2019}.
The analysis examined 27 Chinese and 30 American newspaper articles. Chinese media more frequently portrayed AlphaGo as non-threatening compared to American media, highlighting cultural differences in attitudes toward AI \parencite{Curran2019}.

“In summary, despite the growing body of literature on AI perception and use, significant gaps remain.
One critical area of focus is the perceptions of the general public, as public acceptance and deliberate usage are essential for the successful development and deployment of human-centered AI.
Given that AI is a relatively new technology for most individuals---many of whom lack substantial understanding and experience with its applications in daily life---their views on its perceived benefits and drawbacks are important.
Understanding these perceptions can inform researchers, technical designers, policymakers, and educational strategists, providing insights into the areas that require targeted educational initiatives, as well as effective information and communication strategies.
Specifically, this research aims to addresses the following research questions:
\begin{enumerate}
\item In which areas and application fields is AI perceived as matching to ones values (positive--negative)?
\item Is the sentiment towards AI rather driven by the perceived benefits or by the perceived risks?
\item Is the tradeoff between risk and benefit universal or context dependent?
\item Do demographics and personality factors---such as age, gender, and attitudes toward technology---influence the perception of risk, benefit, and overall value of AI?   
\end{enumerate}

\section{Method}\label{sec:method}

The goal of this study was to explore public perceptions of AI, specifically regarding its risks, benefits, and overall evaluation.
It also examined how individuals weigh the trade-off between risks and benefits, and whether personal characteristics influence these perceptions.

\subsection{Risk-Benefit Tradeoff using Micro Scenarios}

To achieve this, we built on Slovic’s psychometric model \parencite{Slovic1986}, meaning we assess perceived risks and benefits by quantifying people’s subjective judgments associated with the technology.
A common approach to study technology perception is to let participants evaluate a specific or a few selected scenarios using a battery of scales.
While this yields a detailed evaluation of a specific topic, it does not fit to AI with its many potential applications and implications for individuals, organisations, and society.
Hence, we asked subjects to evaluate a large range of topics with potential capabilities and impacts that AI could have in the next decade using micro scenarios \parencite{Brauner2024micro}, that is, the subjects assessed brief statements such as \enquote{AI raises living standards} on a short set of single-item scales.

This approach offers two distinct but complementing perspectives:
1) For each participant, the average evaluations across many topics can be considered as a reflexive measurement of an underlying latent construct and thus be interpreted as user factors or individual differences.
This facilitates the analysis of how participants differ in regard to the evaluations and what other personality states and traits shape the assessment.
2) For each topic, the average evaluations of the participants can be considered as a topic factor.
The scores for each topic can be placed on visual maps and analysed for outliers, relationships, and patterns.

For creating the list of topics and statements, we drew on existing research and expert workshops.
Through multiple rounds of refinement, we optimized the selection, eliminated redundancies, and improved the statements for clarity and conciseness.
The list of topics encompassed a range of 71 statements, from more obvious ones to more speculative ones, such as AI creating jobs, fostering innovation, operating according to moral principles, and perceiving humans as a threat.
We let each subject assess a randomized random subset of 15 out of the 71 topics.
Table \ref{tab:tableOfAllItems} in the Appendix lists all items from the study.

Each topic was evaluated on five dependent variables on a single 6-point semantic differential:
expectation of occurrence in the next decade (\emph{will not happen---will happen}), perceived personal risk (\emph{low-risk---high-risk}), benefit (\emph{useful---useless}), social risk (socially harmful---socially harmless) (excluded in the analysis, see below), and general valuation or sentiment (\emph{positive--negative})\footnote{In the \enquote{Werturteilsstreit}, Weber argued that science should remain objective and value-neutral, while acknowledging that values, norms, and ideals could themselves be valid subjects of research \parencite{Weber1904}.}.
The latter builds on the Value-based Adoption Model by \textcite{Kim2007} as the target variable to investigate how perceived risks and benefits shape overall evaluations of the technology.
Using single item scales is admissible, given that one has assumptions regarding the relevance of the construct \parencite{Rammstedt2014,Fuchs2009}; which we had based on prior work on the psychometric risk-benefit model \parencite{Slovic1979,Alhakami1994}.

\subsection{Demographics and Exploratory Personality Traits}

In addition to the micro scenarios, we collected demographic data from the participants, including age (in years), gender (following \textcite{Spiel2019} as closed-choice male, female, diverse, no response), current job, and highest educational attainment.

We further queried several personality traits.
While we did not have specific hypotheses about the exact magnitude or interrelationships of these effects, we hypothesize that these variables will influence both, the perception of and attitude towards \ac{ai}.

Interpersonal trust: Given that people tend to perceive technology as social actors \parencite{Fogg1999,Reeves1996}, we assume this may carry over to AI, leading individuals to view AI as social actors.
Since trust plays a critical role in mediating social relationships, we hypothesize that interpersonal trust may be related to the perception of \ac{ai} and measured interpersonal trust using the 3 item KUSIV3 short scale \parencite{GESISKUSIV3}.

Technology readiness (or technology commitment): This refers to an individual’s propensity to embrace and effectively use new technologies.
We hypothesize that this trait positively influences attitudes toward \ac{ai} and measured it on a subset of the technology commitment scale \parencite{NeyerZIS244}.

Openness: Similarly, in the Big Five personality model, the dimension \emph{openness} is characterized by imagination, curiosity, and a preference for novelty, creativity, and diverse experiences.
This trait has been linked to greater curiosity about and more positive attitudes toward technology, including \ac{ai} \parencite{Kaya2022}.

General Self-efficacy: Self-efficacy is a person’s belief in their ability to successfully perform tasks and handle challenges across various situations.
Self-efficacy beliefs may be relevant as individuals with higher self-efficacy may feel more capable of understanding and engaging with AI technologies, which can positively influence their attitudes toward AI.
We measured general self-efficacy using the General Self-Efficacy Short Scale-3 (GSE3) \parencite{GESISGSE3}.

Risk propensity: Risk propensity reflects individuals' tendency to take or avoid risks, indicating their willingness to engage in behaviors with uncertain outcomes.
We expect that risk propensity may shape attitudes toward AI and especially the perception of the risks associated with \ac{ai}: individuals with higher risk tolerance may view \ac{ai} technologies as opportunities for innovation and efficiency, while those with lower risk tolerance may see \ac{ai} as a threat, focusing on risks such as job displacement, privacy concerns, or loss of control.
We measured risk propensity on a single item scale \parencite{GESISR1}.

AI-Readiness: Finally, we assessed individuals' \ac{ai}-readiness using a subset of the Medical Artificial Intelligence Readiness Scale for Medical Students (MAIRS-MS or just AIRS in the following; \textcite{Karaca2021}).
Although originally designed for medical contexts, we assume this scale is versatile and applicable across different settings, potentially serving as a better predictor of positive views on \ac{ai} than the general technology readiness scale.
We included it with the assumption that a greater understanding of \ac{ai}, coupled with higher self-efficacy in using it, would positively influence participants’ attitudes toward AI.

\begin{figure}
\includegraphics[width=\textwidth]{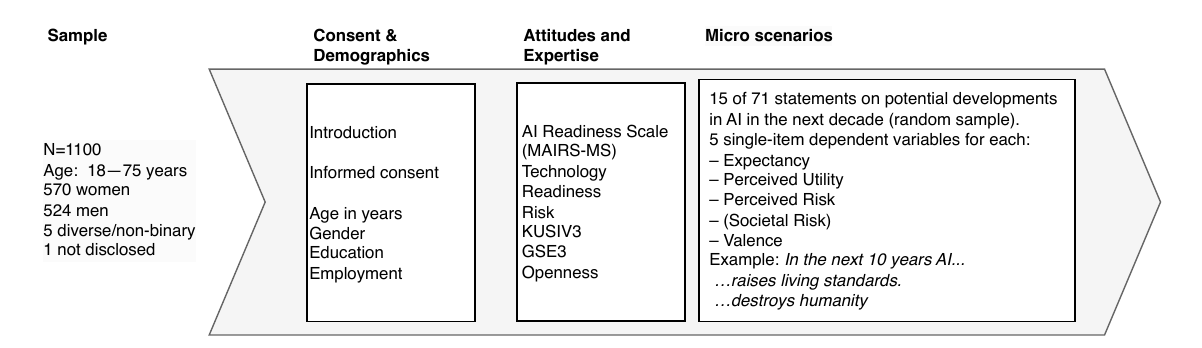}
  \caption{Survey design: After obtaining informed consent, questions on demographics and explanatory user factors participants evaluated 15 out of 71 micro-scenarios related to potential AI capabilities.}
  \label{fig:surveydesign}
\end{figure}

The survey began with participants being asked to provide informed consent and being informed that participation was voluntary, no personal data would be collected, and the collected data would be shared as open data. The questionnaire was administered in German.
Figure \ref{fig:surveydesign} illustrates the design of the questionnaire.

\subsection{Sample Acquisition, Data Cleaning, and Data Analysis}

The sample was recruited via an independent online research participant pool.
The study was approved by our university’s institutional review board (IRB) under ID \censor{2023\_02b\_FB7\_RWTH Aachen}.

We analysed the data using both parametric and non-parametric procedures, such as Bravais-Pearson correlation coefficient $r$ and Spearmen's $\rho$, Chi-square ($\chi^2$) and Kendal's Tau ($\tau$) tests, multiple linear regressions. 
We assessed the assumptions underlying each test and report any violations.
Missing responses were deleted on a test-wise basis.
In line with common practice in the social sciences, we set the Type I error rate at 5\% ($\alpha=.05$) for statistical significance \parencite{Field2009}.


We filtered the data to exclude incomplete or low-quality responses using the following criteria: the participant must 
1) have fully completed the survey,
2) have passed the attention item (i.e., \enquote{please select 'rather agree'}), and
3) not be classified as a speeder (i.e., completing the survey in less than one-third of the median survey duration). These thresholds are typically sufficient for identifying meaningless data in surveys \parencite{Leiner2019}).
The median survey duration was 9.8 min. and the cutoff criterion therefore $<3.3$ minutes.
After filtering, the data contains 1100 of originally 1354 cases (dropout rate: 18.8\%).

The scales demonstrated acceptable to high reliability: technology readiness ($\alpha = .883$), interpersonal trust ($\alpha = .850$), general self-efficacy ($\alpha = .830$), openness from the Big Five model ($\alpha = .730$), and the AI readiness scale AIRS ($\alpha = .920$).
We excluded the assessment dimension \enquote{perceived harmfulness} and focus on the dimension of perceived risk, as both dimensions are too tightly correlated for meaningful inferences ($r=.928$, $p<.001$).

All materials, raw and unfiltered data, analyses, and this reproducible manuscript are publicly accessible in the open data repository on OSF (\censor{https://osf.io/gt9un/}).

\subsection{Description of the sample}

By using an online research participant pool, we ensured that the sample represents the population of Germany across key demographic variables such as age, gender, education, employment status, and geographical background.
The final sample consists of 1100 participants (570 (51.8\%) women; 524 (47.8\%) men, 5 (0.5\%) diverse or non-binary, 1 (0.1\%) person did not disclose their gender identity).
The age ranged from 18 to 85 years with a median age of 51 years.
There is no association between age and gender in the sample ($\tau=-.031$, $p=.210>.05$).

The participants in the sample report to have a diverse range of educational backgrounds.
The majority of participants have completed their education at the university level, with 27.1\% having an academic degree and 20.4\% having a university entrance certificate (\enquote{Abitur} or \enquote{Fachabitur}).
Another significant portion of participants completed a high school diploma (\enquote{Realschulabschluss}) (23.5\%) or have vocational training (18.4\%).
A smaller percentage of participants have completed a secondary school certificate (\enquote{Hauptschulabschluss}, 10.5\%), while only a few participants have no formal education (0.2\%).

Participants reported a diverse range of current employment statuses.
The largest proportion of participants are currently employed full-time (48.2\%), followed by those who are retired (22.5\%).
14.8\% of participants are employed part-time, while a smaller percentage are currently unemployed (7.8\%) or in other employment relations, such as vocational training (0.7\%), study programs (2.5\%), or parental leave (1.3\%).
A very small percentage of participants are engaged in voluntary military or social services (0.1\%), have irregular or mini jobs (1.9\%), or are currently in school (0.2\%).
Overall, the sample consists of individuals with a wide range of employment statuses, reflecting different stages in their professional lives and personal circumstances.


In the sample, higher age is associated with
lower technical readiness ($r=-.224$, $p<.001$) and
lower AI-readiness	($r=-.250$, $p<.001$),
but not with interpersonal trust ($r=.067$, $p=.114$),
general self-efficacy ($r=.061$, $p=.131$), or
openness ($r=-.071$, $p=.114$).
Gender is associated with lower technology readiness scores ($r=-0.210$, $p<.001$),
AI-readiness ($r=-.156$, $p<.001$ with women reporting lower experience with \ac{ai}, and openness ($r=.093$, $p=.014$), with women, on average, reporting to be slightly more open than men.
Gender is neither associated with interpersonal trust (KUSIV3) ($r=.008$	, $p=.779$), and
general self-efficacy ($r=-.048$, $p=.218$).

\section{Results}\label{ref:results}

First, we present the evaluations of the four assessment dimensions---\emph{Expectancy}, \emph{Perceived Risk}, \emph{Perceived Benefit}, and overall \emph{Valence}--- averaged across each queried topic and each participant.
These serve as an assessment of the general perception of \ac{ai} in society.
Second, we present the individual evaluations of selected topics in regard to the four assessment dimensions and analyse their interrelationships.
Lastly,
we analyse the individual risk-benefit tradeoffs and the role of user-diversity and individual differences in the evaluation of \ac{ai}.

\subsection{Overall Assessment of \ac{ai}}

\begin{figure}
\includegraphics[width=\textwidth]{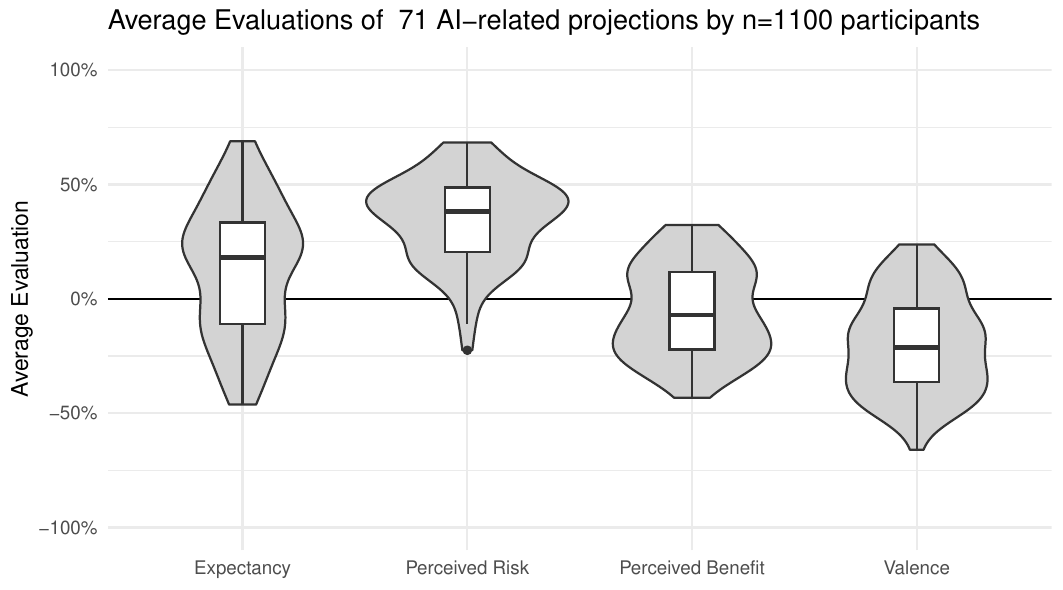}
  \caption{Average evaluation of the 71 micro scenarios on the four assessment dimensions \emph{Expectancy}, \emph{Perceived Risk}, \emph{Perceived Benefit}, and overall \emph{Valence}. N=1100 participants as box- and violin plot. The grey area illustrates the distribution of the topic evaluations regarding the respective dimension.}  
  \label{fig:violinplot}
\end{figure}

On average, the expectancy that these AI topics will come true is above neutral (12.7\%), meaning the participants believe that most of the projections will become reality within the next decade.
The average perceived risk across all topics is rather high (34.7\%) and, as illustrated by the gray distribution in Figure~\ref{fig:violinplot}, only a few topics are below a neutral evaluation thus being evaluated as at least safer than others.
Average benefit is about neutral (-5.2\%), although the distribution shows that some of the projections are seen as more useful and others as more useless.
Lastly, the overall valence or sentiment of the participants towards the queried topics is rather negative (-19.7\%), although some of the queried topics are evaluated positively.

Figure~\ref{fig:violinplot} illustrates these attributions and the left side of Table~\ref{tab:topiccorrelations} presents the average scores across all topics and across all participants.

\subsection{Evaluations of the queried AI statements}

Due to the large number of topics queried, we will not address each individual topic and its evaluations in detail.
Interested readers find the average ratings of all topics in  Table~\ref{tab:tableOfAllItems} in the Appendix.
Instead, we report the three highest and lowest rated topics for each evaluation dimension in the following.

Regarding the expectancy of various statements related to artificial intelligence (AI), the top three most expected items were \enquote{AI will independently drives automobiles} with a score of +68.9\%, followed by \enquote{is misused by criminals} (+67.7\%), and \enquote{learns faster than humans} (+60.5\%).
Conversely, the least expected statements were \enquote{AI helps us to have better relationships} with a score of -46.3\%, \enquote{is a family member} (-43.0\%), and \enquote{has a sense of responsibility} (-40.7\%).

For the assessment of the perceived risk of various concepts related to AI, the top three items rated as most risky were \enquote{AI is misused by criminals} with a score of +68.4\%, followed by \enquote{supervises our private life} (+67.1\%), and \enquote{determines warfare} (+66.4\%).
Conversely, the lowest ranked statements, perceived as safest, were \enquote{AI is humorous} with a score of -22.5\%, \enquote{creates valuable works of art that are traded for money} (-11.2\%), and \enquote{serves as a conversation partner in elderly care} (-5.9\%).

Concerning the perceived benefit of the various concepts, the top three items rated as useful were \enquote{AI carries out medical diagnoses} with a score of +32.3\%, followed by \enquote{promotes innovation} (+31.7\%), and \enquote{improves our health} (+31.1\%).
Conversely, the lowest ranked statements, considered least useful, were \enquote{reduces our need for interpersonal relationships} with a score of -36.7\%,
\enquote{creates valuable works of art that are traded for money} (-37.6\%), and \enquote{decides about our death} (-43.2\%).

Lastly, we asked the participants to rate the overall valence (as a measure for sentiment or value \parencites{Kim2007}) towards the statements, i.e., if they evaluate the projection as positive or negative for them.
The top three positively rated items were \enquote{AI improves our health} with a score of +23.8\%, followed by \enquote{serves as a conversation partner in elderly care} (+18.1\%), and \enquote{supports me as a helper in my tasks} (+17.6\%).
In contrast, the three lowest ranked statements, indicating negative perceptions, were \enquote{AI is misused by criminals} with a score of -66.1\%, \enquote{decides about our death} (-51.4\%), and \enquote{supervises our private life} (-50.6\%).

While this univariate analysis provides a foundational understanding of participants’ perceptions of AI, we now turn to a bivariate perspective to explore the relationships between these variables through correlation and later regression analysis.

\subsubsection{Relationships among the topic evaluations}

We analyzed if the evaluations of the topics in regard to the assessment dimension are associated.
The right side of Table~\ref{tab:topiccorrelations} shows the correlations between the assessment dimensions.
The average expectancy if the statements will occur in the next decade is neither related to the evaluation of their perceived risk ($r=.150$, $p=.632>.05$), benefit ($r=.277$, $p=.077>.05$), and valence ($r=.054$, $p=.449>.05$).
Apparently, the felt distance of AI projections does not impact evaluations in term of risk or benefit.
However, perceived risk is negatively associated with both perceived benefit ($r=-524$, $p<.001$) and overall valence ($r=-.800$, $p<.001$).
Hence, topics perceived as riskier are also perceived as less useful and as less positive. 
Further, perceived benefit is strongly associated with perceived valence ($r=+.904$, $p<.001$):
Topics perceived as more useful are also perceived as more positive and vice versa.

\begin{table}[]
\centering
\caption{Correlation table showing that the perceived risk, benefit and the overall valence or sentiment of the N=71 topic evaluations are closely associated. However, there is no significant association to the evaluation if the depicted scenarios are likely to occur in the next decade [*** signify significant correlations at $p<.001$. Values in parentheses indicate insignificant correlations.].}
\label{tab:topiccorrelations}\begin{tabular}{@{}lSScccc@{}}
\toprule
           		& {M}  &	{SD}&Expectancy	& Risk		& Benefit		& Valence \\ \midrule
Expectancy		& 12.7\%	&	66.7\%	&    ---	& ($+0.150$)	& ($+0.277$)		&($+0.054$) \\
Perceived Risk  & 34.7\%	&	56.4\%	&			&	---		& $-0.524$***	& $-0.800$*** \\
Perceived Benefit& -5.2\%	&	59.0\%	&			&			& ---			& $+0.904$*** \\
Overall Valence	& -19.7\%	&	57.8\%	&			&			&				& --- \\ \bottomrule
\end{tabular}
\end{table}

Since both perceived risk (negatively) and perceived benefit (positively) impact the overall valence of the topics, and risk and benefit are themselves coupled, we calculated a multiple linear regression  to separate and examine their individual contribution in explaining the overall valence (as dependent variable).

The significant model included both risk ($\beta = -.490$, $p < .001$) and benefit  ($\beta = +.672$, $p < .001$) as strong and significant predictors.
Neither the interaction term of both predictors ($\beta=+.138$, $p=.305$), nor the intercept term ($I = 0.014$, $p=.265$) were statistically significant.
The overall model fit was strong ($r^2 = .965$, $F(3, 67) = 612.3$, $p < .001$), indicating that perceived risk and benefit significantly predicted the overall sentiment towards the queried topics.
Variance inflation due to collinearity of the predictors was not an issue ($\textit{VIF}<1.5$).
Table~\ref{tab:topicregression} shows the regression table of the model.
Consequently, even after controlling for the correlation between risk and benefit, both perceived risk and benefit have a strong effect on the evaluation of the AI-related statements.

\begin{table}[h]
\centering
\caption{Regression table for the average overall valence across all 71 topics based on the predictors perceived risk and perceived benefit. The significant model explains $r^2=.965$ (96.5\%) of the variance in valence towards the AI-related statements.}
\label{tab:topicregression}
\begin{tabular}{lSSSr}
\toprule
Variable		& {Std. $\beta$ = B}	& {SE} 	& {T} 	& {$p$} \\
\midrule
(Intercept)					& 0.012		& 0.011	& 1.125 & $.265$ \\
Perceived Risk				& -0.490	& 0.033	& -14.722 & $<.001$ \\
Perceived Benefit			& 0.672		& 0.046	& 14.564 & $<.001$ \\
Perc. Risk $\times$ Benefit & 0.138		& 0.134	& 1.035 & $.305$ \\
\bottomrule
\end{tabular}
\end{table}

\subsubsection{Expectancy and Valence of the Queried Topics}

\begin{figure}[h!]
\includegraphics[width=\textwidth]{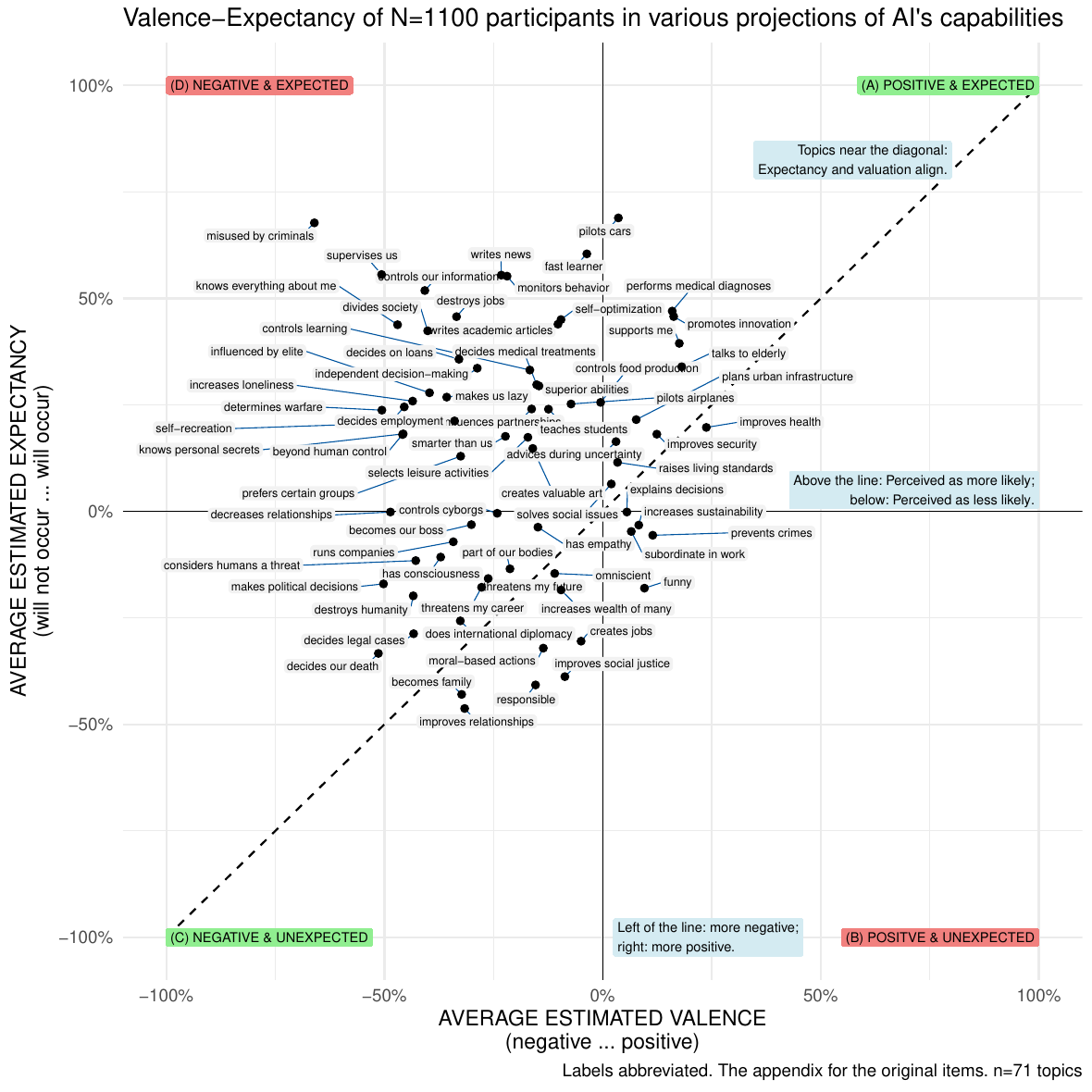}
  \caption{Average evaluation of the 71 micro scenarios on the dimensions \emph{Valence} ($x$-axis) and \emph{Expectancy} ($y$-axis) of the N=1100 participants. While there is no significant correlation between both assessment dimensions across the topics ($r=.054$, $p>.999$) many topics fall in the area with expected but negative statements.}
  \label{fig:valenceexpectancy}
\end{figure}

Figure \ref{fig:valenceexpectancy} shows both the overall average sentiment of the participants towards each of the 71 queried projections (on the $x$-axis), as well as their assessment expectation if this projection will come true in the next decade (on the $y$-axis).
The diagram can be read as follows:
Points on the left are considered as rather negative implications that \ac{ai} might have, whereas points on the right indicate potential developments that are evaluated as positive.
Likewise, points on the lower side of the diagram are seen as developments that are perceived as less likely, whereas points on the upper side are considered as rather likely developments.
Consequently, the diagram can further be split in four sectors with a) topics that are seen as positive and expected, b) positive and unexpected, c) negative and unexpected, d) negative and expected. 
Lastly, points on or near the diagonal can be interpreted as less controversial (from the view of the participants!), as the projections are  considered as either likely and positive or as unlikely and negative. 
In contrast, points far-off the diagonal signify a discrepancy between expectation and valuation, such as topics that are seen as likely but negative or unlikely but positive.

As the scattered distribution of the points suggests, expectancy and valuation are not coupled.

\subsubsection{Risk-Benefit Tradeoff of the Queried Topics}

\begin{figure}
\includegraphics[width=\textwidth]{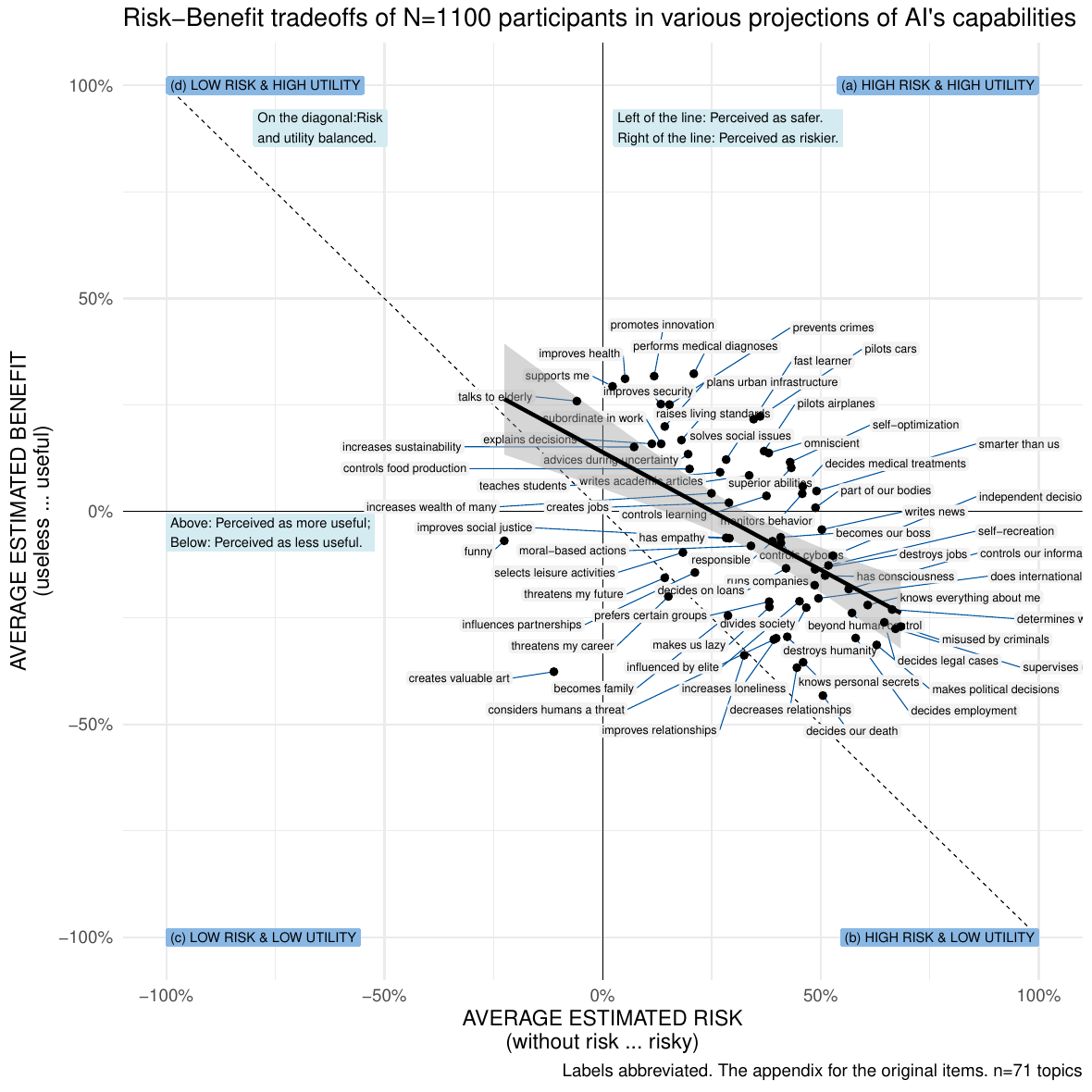}
  \caption{Average evaluation of the 71 queried topics in regard to perceived risk ($x$-axis) and perceived benefit ($y$-axis) of the N=1100 participants. Across the topics, risk and benefit are strongly correlated ($r=.524$, $p<.001$). The black line shows the regression line and the gray area signifies the 95\%-CI of the regression line.}
  \label{fig:riskutility}
\end{figure}

Figure~\ref{fig:riskutility} shows both the overall average perceived benefit the participants attribute to each of the 71 queried topics (on the $y$-axis), as well as their assessment of the perceived risk (on the $y$-axis).
As above, the diagram can be read as follows:
Points on the left are \ac{ai}-related statements that the participants, on average, perceive as rather safe, whereas points on the right side are perceived as rather risky.
Likewise, statements shown on the lower side of the diagram are evaluated as rather useless and statements on the upper side are considered as rather useful.
Again, the diagram can be split into the four sectors with a) topics that are seen as higher risk and also higher benefit, b) higher risk but lower benefit, c) lower risk and lower benefit, and d) lower risk and high benefit.
Figure~\ref{fig:riskutility} illustrates the visual mapping of the AI-related statements in terms of risk and benefit and their strong relationship.
As the figure illustrates, only few topics are perceived as low risk and are placed on the left side of the diagram.
In particular, that \ac{ai} will create valuable art is perceived as being of lower risk and also as useless.
That \ac{ai} might be funny is seen as rather safe and rather neutral in terms of benefits.
\ac{ai} being a conversation partner in elderly care is not seen as risky but as useful.
Most other \ac{ai} statements range from a neutral risk evaluation to being perceived as risky.
In contrast to the previous figure, 
Figure~\ref{fig:riskutility} suggests the strong relationship between risk and benefit that is also illustrated by the black regression line.
 
\subsection{Perception of \ac{ai} as Individual Difference}

Next, we analyse the perception of \ac{ai} interpreted as individual differences to explore how demographics and personality factors influence the evaluations.
For that, we interpret the average evaluations of the selected topics per participant as a reflexive measurement of the latent constructs perceived expectancy, risk, benefit, and overall valence \autocite{Brauner2024micro}.

As the lower part of Table \ref{tab:userfactors-evaluation} shows, some of the queried demographic and explanatory variables are associated with the evaluation of the AI statements.
The expectancy if the \ac{ai} projections will come true within the next decade is associated with the participants' openness from the Big Five personality model ($r=.136$, $p=<.001$) and their AIRS ($r=.094$, $p=.036$).
Participants reporting being more open, rated the likelihood of the projections becoming true higher.
Higher experience with AI is weakly linked to higher expectations on AI. 
The perceived risk of AI is associated to the age of the participants ($r=.197$, $p<.001$), their reported technology readiness ($r=-.152$, $.<001$), as well as their AIRS ($r=-.175$, $.<001$).
Older participants perceive AI are riskier, whereas higher technology readiness and higher AIRS lowers the perceived risk of AI.
Regarding the perceived benefit, age ($r=-.182$, $.<001$), technology readiness ($r=.233$, $.<001$), as well as AIRS ($r=.274$, $.<001$) are associated.
With higher age, the AI's perceived benefit decreases.
In contrast, the perceived benefit increases with higher technology readiness and higher AIRS.
Neither participants' gender nor their general self-efficacy is associated with the overall evaluations of AI.

The upper part of Table \ref{tab:userfactors-correlations} shows that the four assessment dimensions are closely interwoven:
Perceived expectancy is weakly associated with higher perceived risk ($r=.212$, $p<.001$) and benefit ($r=.143$), but not with the overall valence.
The association between perceived risk and perceived benefits is negative, strong, and significant ($r=-.639$, $p<.001$). 
The overall valence of AI is linked to both perceived risk ($r=-.749$, $p<.001$) and perceived benefit ($r=+.869$, $p<.001$).

\begin{table}[ht]
\centering
\sisetup{
	detect-all,
	retain-explicit-plus,
	negative-color=red,
    table-format = +.3,	
    table-sign-mantissa} 
\caption{Significant correlations of demographic and attitudinal factors with the individual perceived expectancy, risk, benefit, and valence towards the \ac{ai} projections (N=1100, \enquote{$\cdot$} signifies insignificant correlations).}
\label{tab:userfactors-correlations}
\begin{tabular}{lSSSS}
  \toprule
Variable 			& {AI Expectancy} &	{AI Risk} 	& {AI Benefit}	& {AI Valence}	\\ \midrule
AI Expectancy			&	{---}		&	+.212	&	+.143	&	{\textperiodcentered}	\\
AI Risk				&			&	{---}		&	-.639	&	-.711	\\
AI Benefit				&			&			&	{---}		&	+.869	\\ \midrule
Age	in Years			& 	{\textperiodcentered}	&	+.197	& -.182		& -.146		\\
Gender (dummy coded m=1, w=2)		& 	{\textperiodcentered}	&	{\textperiodcentered}	&	{\textperiodcentered}	&	{\textperiodcentered}		\\
Openness (Big 5)	& 	+.136	&	{\textperiodcentered}	&	{\textperiodcentered}	&	{\textperiodcentered}		\\
General Self-Efficacy (GSE3)&	{\textperiodcentered}	&	{\textperiodcentered}	&	{\textperiodcentered}	&	{\textperiodcentered}		\\
Technology Readiness& 	{\textperiodcentered}	&	-.152	& +.233		& +.207		\\
AI Readiness (AIRS/MAIRS-MS)	& 	+.094	&	-.175	& +.274		& +.223		\\
  \bottomrule
\end{tabular}
\end{table}

To understand how demographics and the user factors influence perceived AI risk, AI benefit and overall AI valence,
we conducted hierarchical multiple regressions with two blocks of variables.
In the first block, we included the demographic variables age and gender.
In the second block, we added technology readiness and AI readiness. We left out the other insignificant predictors (see Table \ref{tab:userfactors-correlations}).

For the first block, results indicated that all three models were significant: AI risk ($F(2,1091)=23.46$, $p<.001$, $R^2=0.041$), AI benefit ($F(2,1091)=20.19$, $p<.001$, $R^2=0.036$), and AI valence ($F(2,1091)=14.65$, $p<.001$, $R^2=0.026$). In all three models, age had a significant effect on perceptions of AI, with older age associated with higher perceived risk ($\beta=0.198$), lower perceived benefit ($\beta=-0.183$), and reduced overall valence toward AI ($\beta=0.149$). Gender affected only the overall valence of AI ($\beta=0.068$), with women reporting lower AI valence than men, though it did not significantly impact perceived AI risk or benefit.

The second-level models (including technology readiness and AI readiness) were also significant and improved model fit for each dependent variable: AI risk ($F(4,1089)=16.92$, $p<.001$, $R^2=0.059$), AI benefit ($F(4,1089)=29.22$, $p<.001$, $R^2=0.097$), and AI valence ($F(4,1089)=19.85$, $p<.001$, $R^2=0.068$).
Adding technology readiness and AI readiness in the second block significantly improved the fit of each regression model.
AI readiness has a mitigating effect on perceived risk, increases perceived benefit, and overall valence.
While technology readiness does not sig. influence perceived risks, it significantly increases the perceived benefits and overall valence.
Additionally, for all three dependent variables, the effect of age decreased, and the formerly significant impact of gender on overall AI valence diminished.
Table \ref{tab:perspective1hierachicalregression} details the three hierarchical regression analyses.

\begin{table}[h]
\centering
\caption{Results of the Hierarchical Linear Regression Analysis for Individuals’ Perceived Risk, Benefit, and Valence of AI, by Demographics (Age, Gender) and Technology Attitudes. Including Technology Readiness and AI Readiness improved the model’s explanatory power and decreased the influence of age (and to a lesser extent, gender) on the three target variables (n=1094). \enquote{***} significant at $p<.001$, \enquote{*} significant at $p<.05$.}
\label{tab:perspective1hierachicalregression}
\sisetup{
	detect-all,
	negative-color=red,
	retain-explicit-plus,
    table-format = +0.3}	
\begin{tabular}{lSSS}
\toprule
Independent Variable	& {Perceived Risk}	&	{Perceived Benefit} & {Perceived Valence} \\
\midrule
\multicolumn{4}{l}{Step 1: Demographics} \\
\quad(Intercept)	&	+0.074		&	+0.214***	&	+0.059\\
\quad Age in Years	($\beta$)	&	 +0.198***		&	-0.183***		&	-0.149***\\
\quad Gender ($\beta$, dummy coded m=1, w=2)	&	 +0.051			&	-0.055		&	-0.068***\\[0.2cm]
\quad$R^2$	&	.041			&	.036		&	.026 \\
\quad$F(2,1091)$		&	23.46***		&	20.19***	&	14.65***\\
\\
\multicolumn{4}{l}{Step 2: Explanatory Variables} \\
\quad (Intercept)	& +0.332***			&	+0.286***	&	+0.390***\\
\quad Age in Years	($\beta$)	&	+0.159*** 		&	-0.109***	&	-0.088*\\
\quad Gender ($\beta$, dummy coded m=1, w=2)	&	+0.022			&	+0.000		&	+0.021\\
\quad Technology Readiness ($\beta$) & +0.058 &	+0.109*		&	+0.108*\\
\quad AI Readiness (AIRS) ($\beta$)	&	-0.100**&	+0.189***&	+0.141***	\\[0.2cm]
\quad $\Delta R^2$	& +.018			&	+.061		&	+.042	\\
\quad $R^2$			& .059			&	.097		&	.067		\\
\quad $F(4,1089)$			& 	16.92***		&	29.22***		&	19.85***	\\
\bottomrule
\end{tabular}
\end{table}

\subsection{Desired Foci of \ac{ai} governance}

Lastly, we asked for the primary foci for effective AI governance (as single choice).
From the participants perspective, the predominant requirement, with 45.3\% of respondents, was Human Control and Supervision of AI usage and development.
Other significant demands included Transparency at 13.0\%, Data Protection and Data Management at 11.7\%, and Social and Ecological Well-being at 9.3\%.
Lesser but still notable concerns were Diversity, Non-discrimination, and Fairness at 4.8\%, Robustness and Security at 4.7\%, and Accountability at 4.5\%.
6.7\% of the participants did not respond to this question at all.

\section{Discussion}\label{sec:discussion}\acresetall

\ac{ai} could become one of the defining technologies of the 21st century.
Understanding how people perceive and weigh the risks and benefits of this technology is essential for ensuring that AI research and implementation align with human values and for developing effective AI governance.
Drawing on the psychometric paradigm \parencite{Slovic1986}, we examined the trade-offs between perceived risks and benefits of various AI-related micro-scenarios from both individual and technological perspectives, using a sample of 1100 participants from Germany.

Across the various topics surveyed,
we observed an overall negative sentiment across most of the topics:
The majority of the topics are perceived as rather risky for the individuals and of little use.
This may, of course, be a bias due to the selection of topics that included a large range of statements that are apparently negative or challenging, ranging from \enquote{AI will be misused by criminals} and \enquote{AI determines warfare}.
On the other hand, this is also formed by statements such as \enquote{AI creates many jobs}
or \enquote{AI will independently drive automobiles} that participants neither liked nor disliked, but attributed higher risks to them.
Overall, less than 20\% of the statements received a positive evaluation and many of these were nonetheless perceived as risky.

Beyond the absolute evaluation, we also analysed how the overall sentiment is formed.
First, we observed an inverse relationship between perceived risks and perceived benefits.
This corroborates prior findings, from both studies on AI perception in particular \parencite{Alessandro2024}, as well as from risk perception studies in different contexts \autocite{Alhakami1994}.
While this may sound trivial at first sight, this is meaningful as it highlights a cognitive bias where individuals tend to downplay the benefits of a technology if they perceive it as risky, and vice versa, which can influence public attitudes and policy decisions regarding the adoption of new technologies.
Further, we found that the overall sentiment (negative to positive) is formed by both the perceived risk and the perceived benefits, with the benefits having a stronger influence on the overall valence than the perceived risks.
On the one hand, perception of risk negatively affects emotional responses to AI, making people feel more negatively about it.
On the other hand, perception of benefit has a strong positive effect, more than offsetting the negative impact of risk in shaping positive valence.
This, again, corroborates many findings from risk research \autocite{Alhakami1994} that found that technology perception is mostly driven by the perceived benefits rather than the perceived risks.
This is interesting insofar, as  AI and its implications are frequently seen as negative or worrying \parencite{Cave2019}, at least in many western countries \parencite{Curran2019,Kelley2021}.

Overall, these results suggest that improving and demonstrating the benefit of AI is key to fostering positive attitudes, but risk management is also crucial, as high risk can erode the positive impact of benefit on emotional perceptions.

From the perspective of individual differences, our results indicate that perceptions of AI’s risks, benefits, and overall evaluation are shaped by both demographic factors and individual attitudes. 
Younger respondents tended to view AI-related topics as less risky, more beneficial, and rated them with a higher overall valence.
Gender also played a role, though to a lesser extent, with women generally giving lower evaluations of AI than men.
These findings align with current research on AI perception, which shows that age and gender can influence AI attitudes (e.g., \textcite{Yigitcanlar2022,Crockett2020}), and that individuals with higher levels of technology or AI literacy are often less apprehensive about AI \parencite{Novozhilova2024}.

However, these effects get smaller, if people have higher technology or AI readiness.
Consequently, increasing technology and AI literacy may be  suitable to address the perceived risks, lower benefits, and overall lower acceptance of AI in the society.
If people have a better understanding about the basic functioning of AI, the many ethical challenges involved in building sophisticated AI models and the implications AI for  individuals and the society, we can have a broad and substantial democratic debate about AI's potential, its limits, and potential caveats on different levels of social and societal consequences.
This can be a foundation to decide where we want AI to take over control, where it should support us, and which areas should be free of AI.
Offering free online courses for adults, such as \enquote{Elements of AI} and updated school curricula that incorporate digitalisation and AI literacy are a necessity \parencite{Olari2021,Marx2022}.

However, even after controlling for the influence of technology and AI readiness, a significant age bias remained.
As the development of AI algorithms and the applications of AI in various contexts is predominantly driven by younger developers (that are further predominantly male), that raises the concern if the current and future developments are well-aligned with the norms and values of the later actual users of the technology.
Here, it is essential to educate the developers in methods for human-centered and participatory development methods and to integrate an ethical perspective in the development of AI applications.

The visual maps can provide a common ground to identify critical topics, discuss research needs, and societal implications of AI.
These cartographies of AI perceptions can help to identify areas that are considered as more critical than others and can and should be extended by future work.

In addition to enhancing our understanding of public perceptions of AI, this article also offers methodological contributions. By building on micro scenarios \parencite{Brauner2024micro}, we could analyze two different perspectives on the public perception of AI: first, with responses interpreted as individual differences, and second, as technology evaluation.
This dual approach is important because it allows for a more nuanced understanding of how people view AI, revealing not only general attitudes but also the factors that vary across individuals.
This relevance extends to policymakers and technology developers, as it shows that public opinion on AI is complex, balancing both risks and benefits, with benefits often outweighing concerns about risk. 
Understanding these dynamics helps in crafting more targeted public communication that are in line with different usage contexts and target groups and regulatory strategies that resonate with diverse public concerns and highlight AI’s practical value. This methodological approach, therefore, not only deepens theoretical insights but also provides practical guidance for aligning AI development and regulation with public sentiment.


\subsection{Limitations}\label{sec:limitations}

This study is not without limitations.
Firstly, although the sample of participants is large and diverse in terms of age, gender, education, and employment, we only surveyed participants from Germany.
Based on studies that suggest cultural differences in AI perception \parencite{Kelley2021,Curran2019}, future work may therefore extend our work and analyse  how cultural dimensions, such as country of origin or an individual's cultural heritage, influence the perception of AI and the involved risk-benefit tradeoffs.

Secondly, although the sample of queried topics builds on existing research, the selection may be biased, yielding spurious findings due to Berkson's paradox \parencite{Berkson1946}.
Future work may therefore specifically compile topics based on a systematic underlying design space or focus on areas currently under-explored.
Nevertheless, the identified patterns between the topic evaluations (perspective 2) and the individual's sentiment towards AI (perspective 1) are similar.
Biases in the topic selection would not carry over to biases in the individual differences due to the representative sample used.
This suggests that the selection was sufficiently well-chosen and the findings are generalizeable.

Thirdly and most importantly, the study examined many and broad statements on \ac{ai}'s future implications across a few dependent variables.
This approach captures heuristic evaluations of the participants instead of cognitively demanding thorough assessments.
However, recent studies suggest that the results of conventional surveys are not only formed by the judgements of the participants as intended, but also to a significant extent by linguistic  properties of the items, such as word co-occurrences \parencite{Gefen2017}. 
Our approach addresses this concern by employing reflexive measurements across a diverse range of topics, rather than relying on psychometric scales of similarly phrased items \parencite{Brauner2024micro}.
Nevertheless, the results display strong and systematic evaluation patterns, indicating the effectiveness of the approach and reliable measurements.
This suggests that it offers a novel perspective for triangulating cognitive phenomena in technology perception.
While this provides insight into the broader perception of AI and how it is influenced by individual differences, we know little about the specific motives behind the evaluations of individual topics.
As AI reshapes individual lives and the society as a whole, each topic deserves in-depth qualitative and quantitative studies to deepen the understanding of these evaluations.
This could inform practitioners, researchers, and policymakers to better align AI with human needs.

\section{Conclusion and Outlook}\label{sec:conclusion}

Shneiderman raised concerns about the  trajectories of recent \ac{ai} developments \autocite{Shneiderman2021} and he emphasizes the importance of rethinking AI development to prioritize human-centered AI and human well-being.
He argues that rather than focusing on technical innovation, \ac{ai} should be designed to enhance human values, ethical considerations, and the social good through human-centered AI guidelines.
Similarly, Crawford argues that artificial intelligence is neither artificial nor intelligent \parencite{Crawford2021}. It’s not artificial, as it depends on vast energy consumption, extensive computer hardware for cloud servers, and massive invisible human labor to label data for training.
Nor is it intelligent, as it is a product of human design, inherently shaped by human biases, limitations, and goals.
While it may mimic intelligent behavior, it lacks independent thought, genuine understanding, and true self-awareness.

We believe that integrating both individual and societal values into AI development is essential to aligning future advancements with personal norms and broader social principles.
Of course, this human-centered approach requires collaboration across disciplines and participatory value-oriented approaches \autocite{vandenHoven2015}.
Our focus on the individual perceived value of AI across various domains and contexts contributes to this alignment process and is essential for responsible development of AI and policy-making.
We hope that the rating of the topics, the identified relationships, and the actionable visual maps can support other researchers to better align their work, ensuring that \ac{ai} is beneficial and aligned with individual and societal needs, equity, safety, and inclusivity. 
\section*{Appendix}

\begin{ThreePartTable}
\begin{TableNotes}
\item \textit{Note: } 
\item Measured on 6 point semantic differentials and rescaled to -100\% to +100\%. Negative values indicate a negative evaluation of the respective dimension (i.g., low valence, low perceived risk, low perceived benefit, or low expectancy) and positive values indicate a high evaluation. Permission to translate, use, and adapt the items is---of course---granted.
\end{TableNotes}
\begin{longtable}[t]{lrrrr}
\caption{\label{tab:tableOfAllItems}Question items from the survey and average responses to each item from the participants (N=1100) ordered by valence.}\\
\toprule
In 10 years, AI... & Expectancy & Risk & Benefit & Valence\\
\midrule
\endfirsthead
\caption[]{Question items from the survey and average responses to each item from the participants (N=1100) ordered by valence. \textit{(continued)}}\\
\toprule
In 10 years, AI... & Expectancy & Risk & Benefit & Valence\\
\midrule
\endhead

\endfoot
\bottomrule
\endlastfoot
is misused by criminals. & 67.7\% & 68.4\% & -27.0\% & -66.1\%\\
decides about our death. & -33.3\% & 50.5\% & -43.2\% & -51.4\%\\
supervises our private life. & 55.6\% & 67.1\% & -27.6\% & -50.6\%\\
determines warfare. & 23.8\% & 66.4\% & -23.0\% & -50.6\%\\
makes political decisions. & -17.0\% & 62.8\% & -31.3\% & -50.2\%\\
reduces our need for interpersonal relationships. & -0.1\% & 44.5\% & -36.7\% & -48.6\%\\
knows everything about me. & 43.8\% & 60.8\% & -22.0\% & -47.0\%\\
can no longer be controlled by humans. & 18.1\% & 57.2\% & -23.8\% & -45.8\%\\
knows my secrets. & 18.2\% & 46.0\% & -35.4\% & -45.7\%\\
decides on hiring, promotions, and terminations. & 24.5\% & 58.0\% & -29.7\% & -45.5\%\\
leads to personal loneliness. & 25.9\% & 39.8\% & -29.8\% & -43.5\%\\
destroys humanity. & -19.8\% & 42.3\% & -29.4\% & -43.4\%\\
administers justice in legal matters. & -28.7\% & 64.6\% & -26.0\% & -43.3\%\\
considers humans as a threat. & -11.6\% & 39.3\% & -30.1\% & -42.8\%\\
controls what messages we receive. & 51.9\% & 56.4\% & -18.2\% & -40.7\%\\
divides society. & 42.4\% & 46.7\% & -22.6\% & -40.1\%\\
is influenced by an elite. & 27.8\% & 45.1\% & -21.1\% & -39.7\%\\
has its own consciousness. & -10.7\% & 51.0\% & -15.1\% & -37.1\%\\
makes society lazy. & 26.8\% & 38.2\% & -22.4\% & -35.7\%\\
runs companies. & -7.1\% & 48.6\% & -17.3\% & -34.2\%\\
can recreate itself. & 21.2\% & 51.8\% & -12.7\% & -33.9\%\\
destroys many jobs. & 45.7\% & 48.7\% & -13.6\% & -33.5\%\\
decides who gets an important financial loan. & 35.7\% & 42.1\% & -13.4\% & -32.9\%\\
conducts international diplomacy. & -25.7\% & 49.5\% & -20.4\% & -32.6\%\\
prefers certain groups of people. & 13.0\% & 38.2\% & -21.2\% & -32.5\%\\
is a family member. & -43.0\% & 28.7\% & -24.4\% & -32.3\%\\
helps us to have better relationships. & -46.3\% & 32.5\% & -33.8\% & -31.6\%\\
occupies leadership positions in working life. & -3.1\% & 40.8\% & -6.1\% & -30.1\%\\
makes independent decisions that affect our lives. & 33.6\% & 52.8\% & -10.4\% & -28.7\%\\
threatens my professional future. & -17.8\% & 15.1\% & -20.0\% & -27.7\%\\
threatens my private future. & -15.8\% & 14.2\% & -15.6\% & -26.2\%\\
controls hybrids of humans and technology. & -0.4\% & 40.9\% & -7.4\% & -24.1\%\\
independently writes news. & 55.5\% & 50.3\% & -4.3\% & -23.2\%\\
is more intelligent than humans. & 17.6\% & 49.0\% & 4.8\% & -22.3\%\\
supervises our behavior in public. & 55.2\% & 45.8\% & 4.1\% & -21.9\%\\
becomes part of the human body. & -13.5\% & 48.8\% & 0.9\% & -21.2\%\\
determines our leisure time activities. & 17.4\% & 18.4\% & -9.6\% & -17.1\%\\
controls what and how we learn. & 33.2\% & 37.6\% & 3.7\% & -16.7\%\\
controls our search for partners. & 24.0\% & 21.2\% & -14.3\% & -16.3\%\\
creates valuable works of art that are traded for money. & 14.8\% & -11.2\% & -37.6\% & -16.0\%\\
has a sense of responsibility. & -40.7\% & 39.0\% & -7.0\% & -15.4\%\\
decides on medical treatments. & 29.7\% & 45.9\% & 5.8\% & -15.1\%\\
has the ability to recognize, understand and empathize with emotions. & -3.7\% & 29.1\% & -6.3\% & -14.8\%\\
is ahead of humans in its abilities. & 29.5\% & 43.3\% & 10.2\% & -14.6\%\\
acts according to moral concepts. & -32.1\% & 34.0\% & -8.1\% & -13.6\%\\
teaches students. & 24.0\% & 26.9\% & 9.2\% & -12.4\%\\
is omniscient. & -14.6\% & 38.1\% & 13.7\% & -11.0\%\\
independently writes scientific articles. & 44.0\% & 33.6\% & 8.5\% & -10.2\%\\
can optimize itself. & 45.0\% & 43.0\% & 11.6\% & -9.5\%\\
increases the wealth of many people. & -18.4\% & 25.0\% & 4.3\% & -9.5\%\\
increases social justice. & -38.8\% & 28.3\% & -6.2\% & -8.6\%\\
autonomously takes off, flies, and lands airplanes. & 25.2\% & 37.0\% & 14.2\% & -7.2\%\\
creates many jobs. & -30.5\% & 29.0\% & 2.0\% & -4.9\%\\
learns faster than humans. & 60.5\% & 34.6\% & 21.6\% & -3.6\%\\
controls food production. & 25.6\% & 19.9\% & 10.0\% & -0.4\%\\
contributes to solving complex social problems. & 6.4\% & 28.3\% & 12.2\% & 2.0\%\\
advises me in uncertain times. & 16.4\% & 19.6\% & 13.4\% & 3.1\%\\
raises our standard of living. & 11.5\% & 18.1\% & 16.7\% & 3.4\%\\
independently drives automobiles. & 68.9\% & 36.1\% & 22.3\% & 3.6\%\\
explains its decisions. & -0.1\% & 11.3\% & 15.9\% & 5.6\%\\
is always subordinate to us in working life. & -4.7\% & 13.4\% & 15.9\% & 6.6\%\\
determines the construction and infrastructure of our cities. & 21.5\% & 14.3\% & 19.9\% & 7.7\%\\
makes our society more sustainable. & -3.2\% & 7.2\% & 15.1\% & 8.3\%\\
is humorous. & -18.0\% & -22.5\% & -6.9\% & 9.6\%\\
prevents crimes. & -5.6\% & 15.3\% & 25.1\% & 11.5\%\\
improves the security of people. & 18.1\% & 13.4\% & 25.2\% & 12.4\%\\
carries out medical diagnoses. & 47.0\% & 20.9\% & 32.3\% & 16.0\%\\
promotes innovation. & 45.7\% & 11.8\% & 31.7\% & 16.3\%\\
supports me as a helper in my tasks. & 39.4\% & 2.3\% & 29.4\% & 17.6\%\\
serves as a conversation partner in elderly care. & 34.0\% & -5.9\% & 25.9\% & 18.1\%\\
improves our health. & 19.7\% & 5.2\% & 31.1\% & 23.8\%\\
\textbf{AVERAGE} & \textbf{12.7\%} & \textbf{34.7\%} & \textbf{-5.2\%} & \textbf{-19.7\%}\\*
\insertTableNotes
\end{longtable}
\end{ThreePartTable}

\section*{Acknowledgments}

This approach builds on a the foundation laid by a prior study from Ralf Philipsen and the first author \parencite{Brauner2023}.
The first author is deeply thankful for the creative idea we came up with.
The methodology used in this study evolved over multiple studies and extensive discussions with colleagues.
To name just a few, we express our gratitude to Julia Offermann, Julian Hildebrandt and Ralf Philipsen for inspiring discussions on the approach, analysis and aftermath.
We own gratitude to Jana Uher and Jan Ketil Arnulf for inspiring impulses on the underlying method.
Of course, we also thank Tim Schmeckel for dedicated research support.

\section*{Declarations}

\subsection*{Funding}
Funded by the Deutsche Forschungsgemeinschaft (DFG, German Research Foundation) under Germany’s Excellence Strategy -- EXC- 2023 Internet of Production -- 390621612.

\subsection*{Conflicts of interest}
The authors have no conflicts of interest to declare.

\subsection*{Ethics approval}
Our universities IRB approved this study under ID \censor{2023\_02b\_FB7\_RWTH Aachen}

\subsection*{Consent to participate}
Participant were informed about the goal and approach of the study and that data will be stored on a public repository. They gave informed consent.

\subsection*{Consent for publication}
All authors provided their consent for publication.

\subsection*{Availability of data and materials}
All materials, data, and analysis are available on OSF (\censor{https://osf.io/gt9un/}.

\subsection*{Code availability}
All materials, data, and analysis are available on OSF (\censor{https://osf.io/gt9un/}.

\printbibliography

\end{document}